\documentclass{PoS}

\title{Thermodynamics of strongly interacting plasma with high accuracy}

\ShortTitle{Thermodynamics of strongly interacting plasma with high accuracy}

\author{Leonardo Giusti\\
  Theoretical Physics Department, CERN, Geneva, Switzerland\\
  and
        Dipartimento di Fisica, Universit\`a di Milano-Bicocca\\
        and 
INFN, sezione di Milano-Bicocca\\
        Edificio U2, Piazza della Scienza 3\\ 
        20126 Milano, Italy.\\
       E-mail: \email{Leonardo.Giusti@mib.infn.it}}

\author{\speaker{Michele Pepe}\\
        INFN, Sezione di Milano-Bicocca\\ 
        Edificio U2, Piazza della Scienza 3\\ 
        20126 Milano, Italy.\\
        E-mail: \email{Michele.Pepe@mib.infn.it}}

\abstract{The equation of state of $SU(3)$ Yang-Mills theory is investigated in the
  framework of a moving reference frame. Results for the entropy density, the pressure,
  the energy density, and the trace anomaly are presented for temperatures ranging from 0
  to 230 $T_c$, with $T_c$ the deconfinement temperature. The entropy density is the
  primary observable that has been measured and form which the other thermodynamic
  quantities are obtained. At least 4 different values of the lattice spacing have been
  considered at each physical temperature in order to extrapolate to the continuum limit. 
  The final accuracy is 0.5\%, increasing to about 1\% close to the phase
  transition. A detailed comparison with the results available in the literature is discussed.
}

\FullConference{34th annual International Symposium on Lattice Field Theory\\
		24-30 July 2016\\
		University of Southampton, UK}

\begin{document}

\section{Introduction}\label{intro}

The collective behavior of a plasma of strongly interacting particles has determined the
evolution of the Universe in its early stages. Those extreme conditions can be now reproduced
and investigated at the colliders of heavy ions. Thus, it is highly interesting to interpret
the experimental results with the theoretical predictions coming from Quantum Chromodynamics (QCD).
The lattice regularization of QCD allows to perform first principle calculations of
physical observables by Monte Carlo simulations. Indeed, many numerical studies have
been done to measure the thermodynamic features of QCD at equilibrium and, in particular,
the Equation of State. 

Those calculations are demanding from the computational viewpoint and during the
last 20 years the accuracy of the numerical data has greatly improved as well as the range
of temperatures that has been explored. A first accurate investigation of the Equation of
State in $SU(3)$ Yang-Mills theory was presented in~\cite{Boyd:1996bx}. 
The main thermodynamics quantities -- pressure $p$, entropy density $s$ and energy
density $\epsilon$ -- were measured up to temperatures $T/T_c \simeq 5$, where $T_c$ is
the deconfinement temperature. The method that was proposed in~\cite{Boyd:1996bx} has, by
now, become the standard technique used in many numerical investigations of thermal quantum field
theories. It is based on the direct measurement of the trace anomaly, $(\epsilon- 3  p )$, of the
energy-momentum tensor $T_{\mu\nu}$ in Monte Carlo simulations; the pressure is then
obtained by integrating in the temperature while the entropy density and the energy
density are calculated using the thermodynamic relation $s=(\epsilon + p )/T$. 

In~\cite{Borsanyi:2012ve}, the Equation of State was computed in a broad range of
temperatures, from 0 up to $T/T_c\sim 1000$ with a permille accuracy. A Symanzik
improved action~\cite{Curci:1983an,Luscher:1985zq} has been considered together with a
refinement of the method of~\cite{Boyd:1996bx}. The numerical data show a significative
discrepancy with the results of~\cite{Boyd:1996bx} in a region near $T_c$. 

The approach used in~\cite{Boyd:1996bx,Borsanyi:2012ve} is very effective but it has the drawback
that an ultraviolet power divergence has to be removed by a subtraction at $T = 0$, or at
some other temperature. An alternative method not affected by
ultraviolet power divergences has been exploited in~\cite{Giusti:2014ila,GiustiPepe2016}.
It relies on the formulation of a thermal quantum field theory in a moving reference
frame~\cite{Giusti:2012yj,Giusti:2011kt,Giusti:2010bb} where the entropy density is directly related
to the expectation value of the space-time component, $\langle T_{0k} \rangle$, of the
energy momentum tensor~\cite{Landau1987}. Another approach based on the Gradient
Flow~\cite{Luscher:2010iy} has been also recently explored~\cite{Kitazawa:2016dsl}. 

The renormalization of the bare energy-momentum tensor on the
lattice~\cite{Caracciolo:1989pt,Caracciolo:1991cp} is an important step for obtaining the
thermodynamic features of a thermal quantum field theory from Monte Carlo simulations. A
non-perturbative definition and computation of the renormalization factors of the
energy-momentum tensor have been discussed in~\cite{Giusti:2015daa}.  

Interestingly, once the renormalization factor, $Z_T$, of the off-diagonal component of the $T_{\mu\nu}$ is known,
the entropy density can be easily computed in the framework of a moving reference frame. 
The appealing part of this approach is that the calculation can be performed in a fully
independent way at every temperature. Moreover, since the entropy density is a physical
quantity, the continuum limit is attained by simply changing the bare coupling and the
temporal extent of the lattice so to keep fixed the temperature in physical units.

In~\cite{GiustiPepe2016} we present an accurate determination of the temperature dependence of the
entropy density in $SU(3)$ Yang-Mills theory using the method above. Results for the
pressure, the energy density and the trace anomaly are discussed as well. In these proceedings
we focus on a detailed comparison of our results with those available in the
literature. In particular, we consider the data presented in~\cite{Boyd:1996bx,Borsanyi:2012ve}.

\section{Thermodynamics in a moving frame}

We consider the thermal $SU(3)$ Yang-Mills theory in the Euclidean space in a
moving reference frame. The theory is regularized on a lattice of size $L_0\times L^3$ and
spacing $a$; the spatial vector ${\bf \xi}$ characterizes the moving frame~\cite{Giusti:2012yj,Giusti:2011kt,Giusti:2010bb}. 
The gauge field, $U_\mu (x) \in SU(3)$, is defined on the links and it satisfies the following shifted boundary
condition

\begin{equation}\label{Ashift}
U_\mu (L_0,{\bf x}) = U_\mu (0,{\bf x} - L_0 {\bf \xi})
\end{equation}
along the temporal direction; periodic boundary conditions are set in space. The
dynamics of the theory is described by the Wilson action

\begin{equation}\label{Wilsonact}
S[U] = \frac{3}{g_0^2}\, \sum_{x} \sum_{\mu,\nu} 
\left[1 - \frac{1}{3} \mbox{Re}\, \mbox{Tr} \Big\{U_{\mu\nu}(x)\Big\}\right]
\end{equation} 
where $g_0^2$ is the bare gauge coupling and $U_{\mu\nu}$ is the plaquette field

\begin{equation}\label{plaquette}
U_{\mu\nu}(x) = 
U_\mu (x) U_\nu(x+\hat\mu) U_\mu^\dagger(x+\hat\nu) U_\nu^\dagger (x).
\end{equation} 
The energy-momentum tensor $T_{\mu\nu}$ is defined by

\begin{equation}\label{Tmunu}
T_{\mu\nu} (x) = \frac{1}{g_0^2} \left[
F_{\mu\rho} ^a (x) F_{\nu\rho} ^a (x)  -\frac{1}{4} \delta_{\mu\nu} F_{\mu\rho} ^a (x) F_{\nu\rho} ^a (x) \right]
\end{equation}
in terms of the field strength on the lattice

\begin{equation}\label{Fmunulatt}
F^a_{\mu\nu}(x) = - \frac{i}{4 a^2} 
\mbox{Tr} \Big\{\Big[Q_{\mu\nu}(x) - Q_{\nu\mu}(x)\Big]T^a\Big\}\; , 
\end{equation}
where $Q_{\mu\nu}(x) = U_{\mu\nu}(x) + U_{\nu-\mu}(x) + U_{-\mu-\nu}(x) + U_{-\nu\mu}(x)$,
with the minus sign standing for the negative orientation. 

In the framework of shifted boundary conditions, the off-diagonal components of
$T_{\mu\nu}$ may pick up non-vanishing expectation values. In particular, the entropy density
$s(T)$ at temperature $T^{-1}=L_0\sqrt{1+{\bf \xi}^2}$ is related to the expectation value of the space-time
component of the energy-momentum tensor by the equation

\begin{equation}\label{entropy}
\frac{s(T)}{T^3}= \frac{L_0^4 (1+{\bf \xi}^2)^3}{\xi_k} \, \langle T_{0k} \rangle_\xi \; Z_T.
\end{equation}
Once the renormalization factor $Z_T (g_0^2)$  of the space-time components $T_{0k}$ is determined,
eq.~(\ref{entropy}) provides a simple method for measuring the entropy density. A single
Monte Carlo simulation for measuring $\langle T_{0k} \rangle_\xi$ is required and the
continuum limit is attained in a simple way by increasing $L_0$ and tuning $g_0^2$ so that
the temperature stays unchanged in physical units. We have used the calculation of
$Z_T(g_0^2)$ presented in~\cite{Giusti:2015daa,GiustiPepe2016}. 

\section{The numerical study}
In this section we present and discuss the results of the numerical study to measure the
Equation of State of the $SU(3)$ Yang-Mills theory. Monte Carlo simulations with shifted
boundary conditions have been done to compute the expectation value $\langle T_{0k}
\rangle_{\bf\xi}$. In this calculation we have almost always
considered the shift ${\bf \xi} = (1,0,0)$ since very small lattice artifacts have been
previously observed for that case~\cite{Giusti:2012yj,Giusti:2014ila,Giusti:2015daa}. Only for the
temperature $T/T_c=0.980$, the shift ${\bf \xi} = (1,1,0)$ has been taken into account
while ${\bf \xi} = (1,1,1)$ has been used for $T/T_c=$ 0.904, 1.061 and 2.30.

An accurate computation of $\langle T_{0k}  \rangle_{\bf\xi}$ is not difficult since it is a
well-behaved and ultralocal observable. Moreover, numerical simulations on systems with large
spatial size are not more demanding since the additional cost of updating a large lattice is
exactly compensated by the increased statistics. For the continuum limit extrapolation, we
have carried out numerical simulations with $L_0/a=5$, 6, 7, 8 and, sometimes, also $L_0=3$,
4 and 10. The lattice size in the spatial directions has been chosen to be $L=128$ for
$L_0$ up to 6 and $L=256$ for larger values. Thus, finite size effects are smaller than the numerical accuracy.

The values of the coupling constant $g_0^2$ have been determined as in~\cite{Giusti:2014ila}. 
For temperatures up to $2\, T_c$, we have used the data for the Sommer scale $r_0/a$
in~\cite{Necco:2001xg}; for higher temperatures, we have performed a quadratic
interpolation in $\ln (L/a)$ of the data for the Schr\"odinger functional coupling
constant, $\overline{g}^2(L)$, listed in
Tables A.1 and A.4 of~\cite{Capitani:1998mq}. The critical temperature in units of the Sommer scale is
$r_0 T_c = 0.750(4)$ \cite{Boyd:1996bx,Lucini:2003zr}.

The lattice artifacts turn out to be small with the shifts
${\bf \xi} = (1,1,0)$ and ${\bf \xi} = (1,1,1)$ and even barely visible for ${\bf \xi} =
(1,0,0)$. We have extrapolated the data linearly in $(a/L_0)^2$; when lattice artifacts
could not be observed, a constant fit has been considered. We point out that most
of the uncertainty in $s/T^3$ comes from the renormalization factor: in fact, 
$\langle T_{0k} \rangle_{\bf\xi}$ can be easily measured with an accuracy of 1 permille or better.

\subsection{The entropy density} 
The entropy density $s(T)/T^3$ is the primary observable that we have measured. In order
to calculate the Equation of State, 35 values of the temperature $T/T_c$ in the range
between 0.66 and 230 have been considered. One third of the data (12) has been collected above
$5\, T_c$, where the temperature dependence is very mild, the other ones (23) have been
useful for having a precise determination of $s(T)/T^3$ in the region where the change in
the temperature is stronger. 

\begin{figure}[htb]
\centering
\begin{minipage}{.5\textwidth}
  \centering
  \includegraphics[width=1.\textwidth,height=6cm]{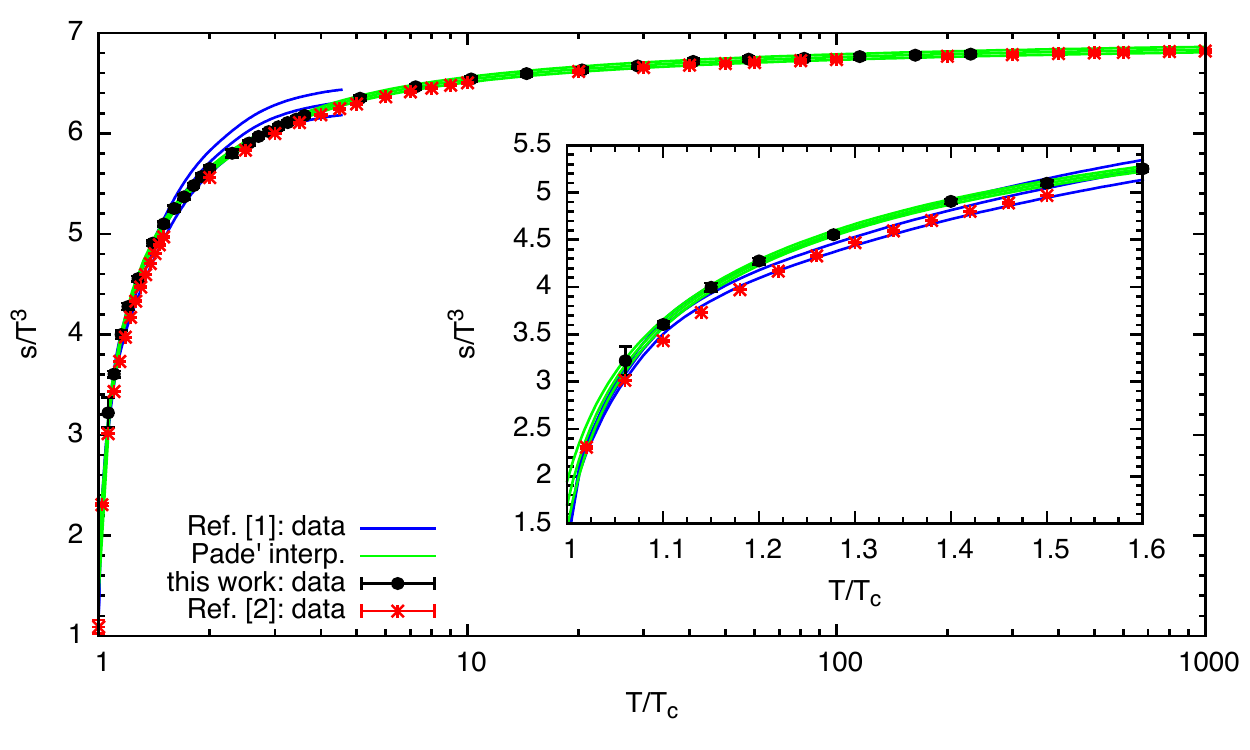}
  \label{figsT3:1}
\end{minipage}%
\begin{minipage}{.5\textwidth}
  \centering
  \includegraphics[width=1.\textwidth,height=6cm]{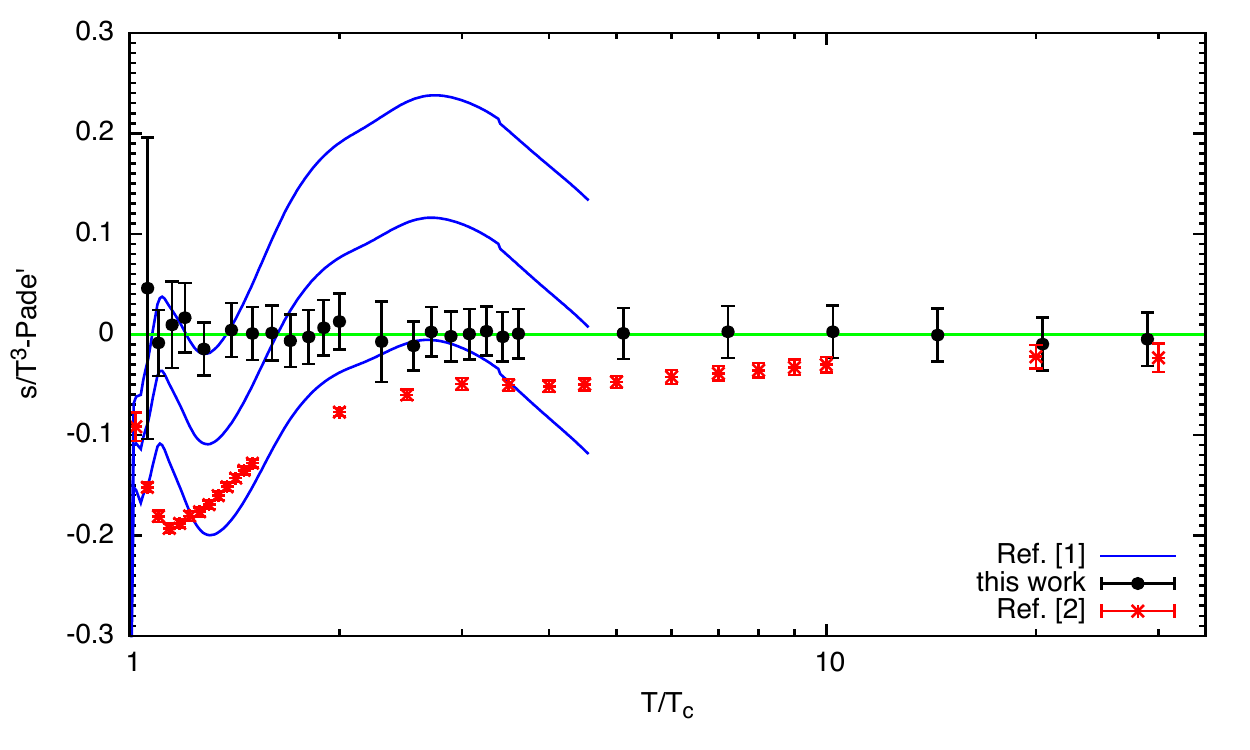}
  \label{figsT3:2}
\end{minipage}
\caption{Left: comparison of the results for the Equation of State of $s(T)/T^3$ with data
  available in the literature. The green band describes a Pad\'e interpolation of our
  numerical data. Right: zoom of the data shifted by the Pad\'e interpolation.\label{sT3}} 
\end{figure}

In the left panel of figure~\ref{sT3} we plot our results together with those available
in~\cite{Boyd:1996bx,Borsanyi:2012ve}. Although one can observe a qualitative agreement between the data,
at a closer look there is a significative and systematic disagreement with the data
published in~\cite{Borsanyi:2012ve}. In the right panel of figure~\ref{sT3}, we show the
data shifted by the Pad\'e interpolation of our numerical data. The discrepancy amounts up
to about 6 standard deviation for $T/T_c$ around 1.24. As the temperature increases, the
discrepancy progressively reduces; for $T/T_c$ larger than 10, the results agree within
error bars. Our data agree with~\cite{Boyd:1996bx} within error bars.

\subsection{The pressure}
The precise determination of the entropy density allows to calculate the pressure by
using the thermodynamic formula $dp(T)= s(T)\, dT$. We have integrated $s(T)$ in $T$ and
we have obtained a precise determination of the temperature dependence of the
pressure~\cite{GiustiPepe2016}. The accuracy and the fine graining of the entropy density
data show that the arbitrariness in the data interpolation necessary to obtain the pressure by
integration, is well below the statistical uncertainty. 

\begin{figure}[htb]
\centering
\begin{minipage}{.5\textwidth}
  \centering
  \includegraphics[width=1.\textwidth,height=6cm]{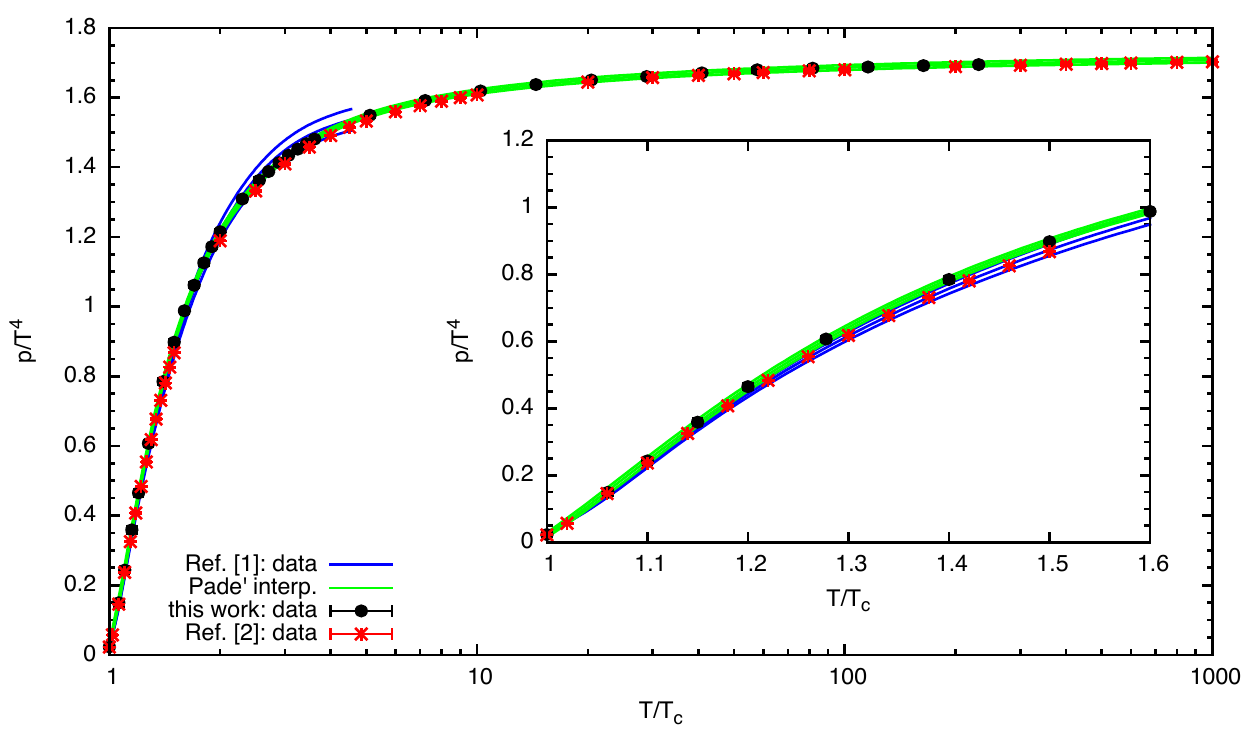}
  \label{figpT4:1}
\end{minipage}%
\begin{minipage}{.5\textwidth}
  \centering
  \includegraphics[width=1.\textwidth,height=6cm]{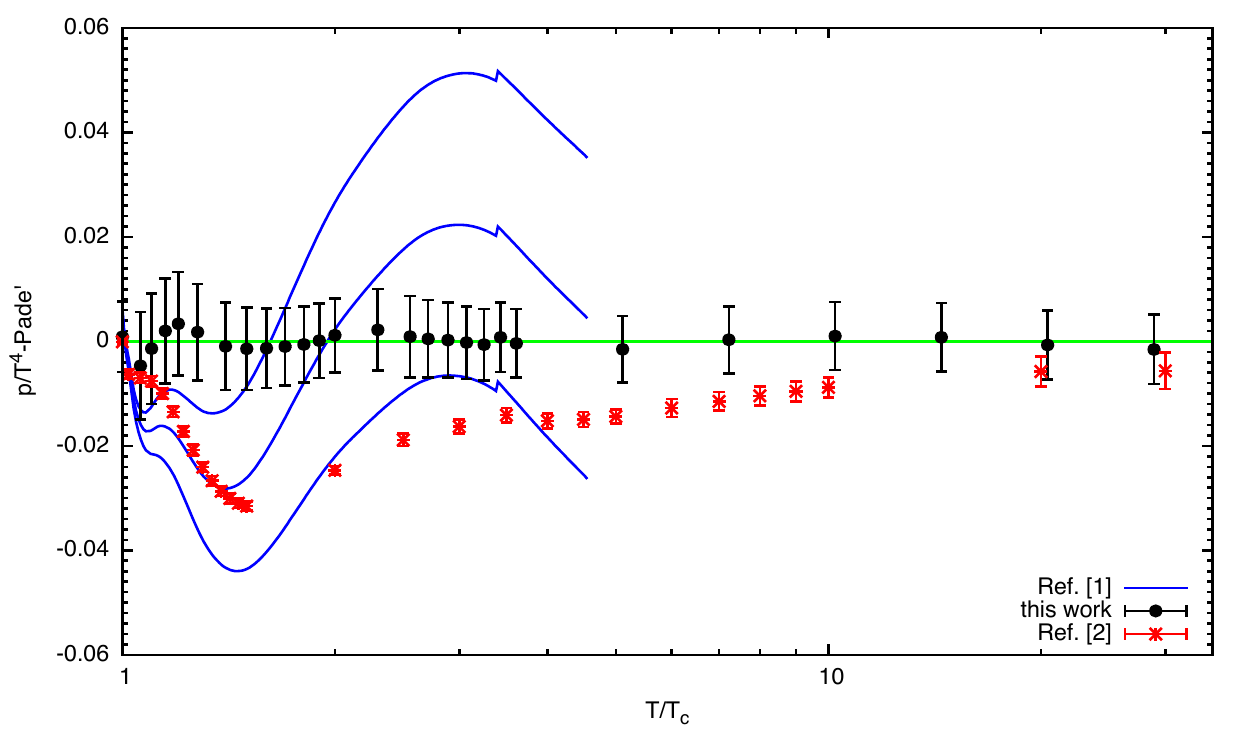}
  \label{figpT4:2}
\end{minipage}
\caption{Left: comparison of the results for the Equation of State of $p(T)/T^4$ with data
  available in the literature. The green band describes a Pad\'e interpolation of our
  numerical data. Right: zoom of the data shifted by the Pad\'e interpolation.\label{pT4}} 
\end{figure}

In the left panel of figure~\ref{pT4} we plot our results together with those available
in~\cite{Boyd:1996bx,Borsanyi:2012ve}. The discrepancy observed in $s(T)/T^3$ with~\cite{Borsanyi:2012ve}, induces a
similar significative discrepancy also in the pressure. This can be seen in the right
panel of figure~\ref{pT4} where the data are shifted by the Pad\'e interpolation of our data.
The disagreement amounts up to 4 standard deviation for $T/T_c$ around 1.5. Again, as the
temperature increases, the discrepancy progressively reduces and, for $T/T_c$ larger than
about 10, the results agree within error bars. Our results are consistent with~\cite{Boyd:1996bx}.

\subsection{The energy density}
\begin{figure}[htb]
\centering
\begin{minipage}{.5\textwidth}
  \centering
  \includegraphics[width=1.\textwidth,height=6cm]{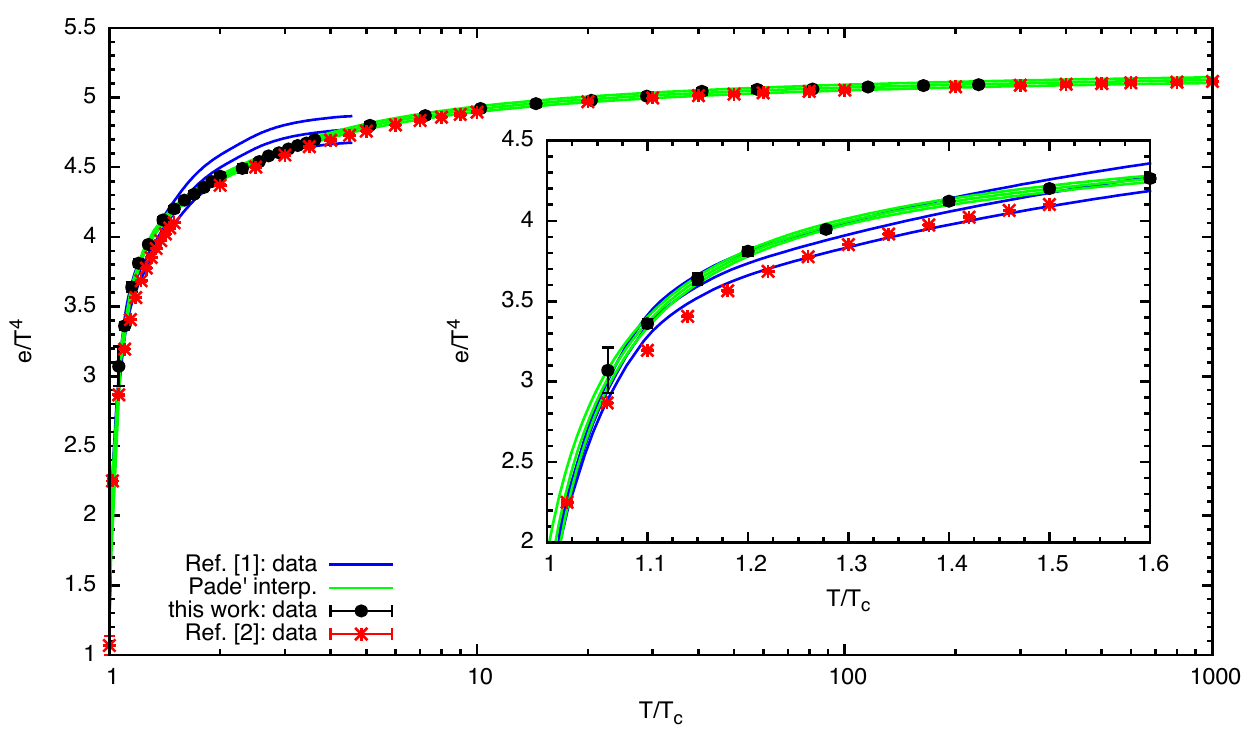}
  \label{figeT4:1}
\end{minipage}%
\begin{minipage}{.5\textwidth}
  \centering
  \includegraphics[width=1.\textwidth,height=6cm]{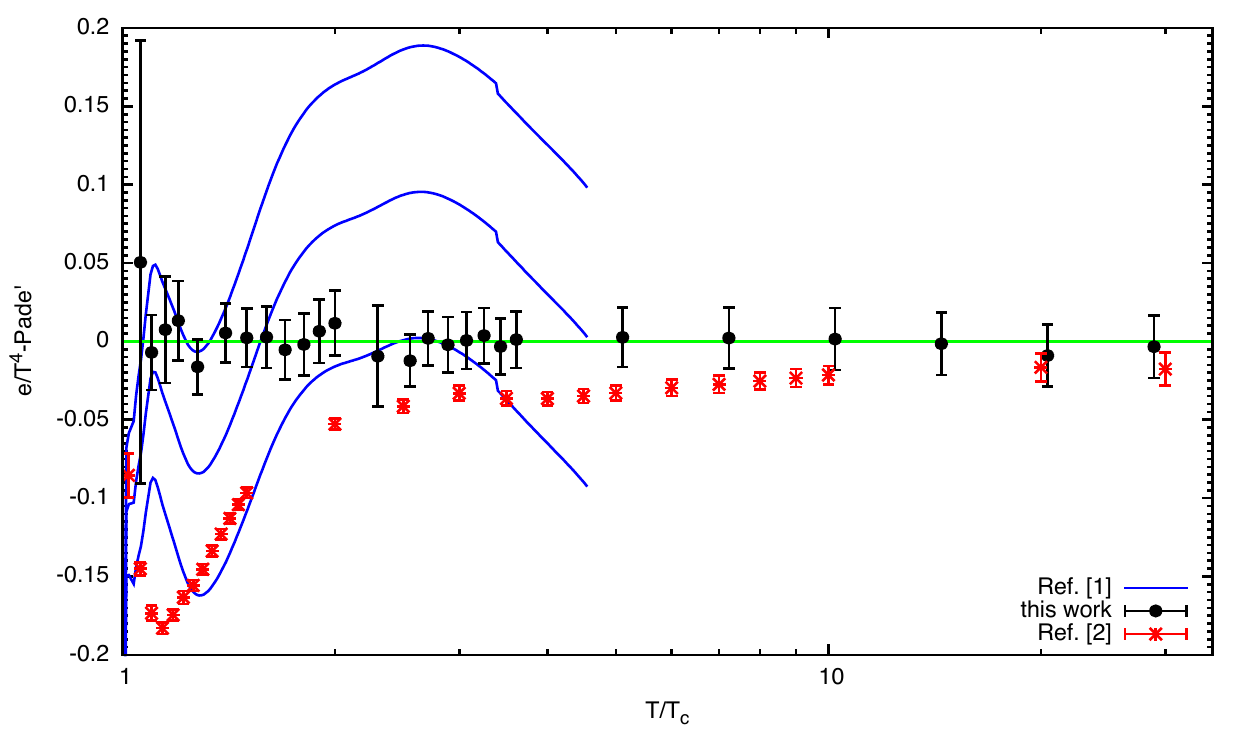}
  \label{figeT4:2}
\end{minipage}
\caption{Left: comparison of the results for the Equation of State of $e(T)/T^4$ with data
  available in the literature. The green band describes a Pad\'e interpolation of our
  numerical data. Right: zoom of the data shifted by the Pad\'e interpolation.\label{eT4}} 
\end{figure}

The calculation of the entropy density and of the pressure allows to obtain the energy
density using the algebraic expression $\epsilon = T s - p$. Another equivalent
option is to integrate the equation $d\epsilon = T ds$. In the left panel of figure~\ref{eT4} we plot our
results together with those published in~\cite{Boyd:1996bx,Borsanyi:2012ve}. In the right
panel, data are shifted by the Pad\'e interpolation of our data. The disagreement
with~\cite{Borsanyi:2012ve} amounts up to 8 standard deviation for $T/T_c$ around
1.22. As for the entropy density and for the pressure, the discrepancy progressively
vanishes when the temperature goes large. Our results agree with~\cite{Boyd:1996bx}.

\subsection{The trace anomaly}
\begin{figure}[htb]
\centering
\includegraphics[width=.8\textwidth,height=6.5cm]{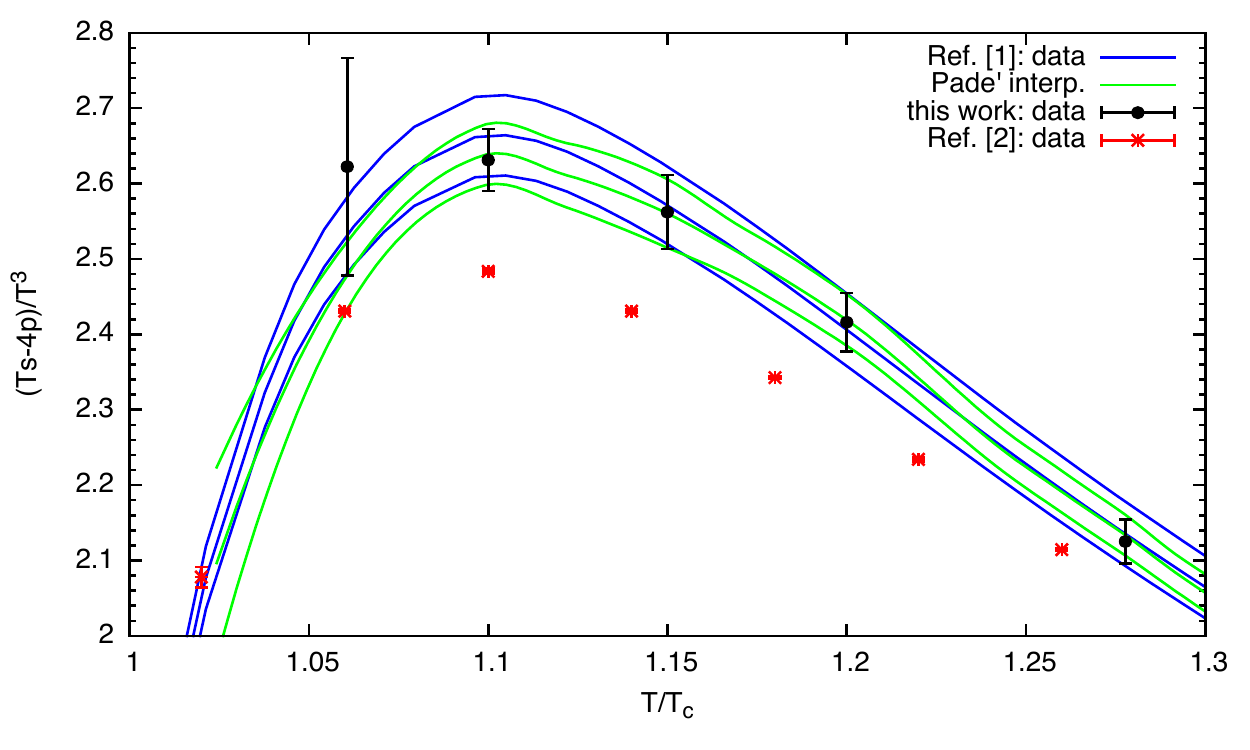}
\caption{Comparison with data in~\cite{Boyd:1996bx,Borsanyi:2012ve} for the trace anomaly
  $(Ts -4\, p)/T^4$ in the region of the peak.\label{anoT4}} 
\end{figure}

The last quantity that we consider is the trace anomaly. In figure~\ref{anoT4} we observe a systematic deviation of
our data for the trace anomaly w.r.t. those presented in~\cite{Borsanyi:2012ve} and an agreement within the numerical accuracy
with~\cite{Boyd:1996bx}. Discrepancies with~\cite{Borsanyi:2012ve} can also be observed in~\cite{Giusti:2014ila,Umeda:2014ula,Kitazawa:2016dsl}.
Interestingly, our data in the continuum limit are in fairly good
agreement with those obtained in~\cite{Borsanyi:2012ve} at their finest lattice spacing.

In the approach used in~\cite{Boyd:1996bx,Borsanyi:2012ve}, the trace anomaly is the
primary observable measured in Monte Carlo simulations and the other thermodynamic
quantities are derived from it. In particular, the pressure is obtained by integrating in
the temperature and energy density and entropy density are derived by algebraic combinations with
the trace anomaly. We expect that the observed discrepancy close to the peak of the trace
anomaly propagates its effects to the larger temperatures where our data agree
within error bars with those in~\cite{Borsanyi:2012ve}.

\noindent {\bf Acknowledgments}
Simulations have been performed on the BG/Q Fermi and on the
PC-cluster Galileo at CINECA (CINECA-INFN and CINECA-Bicocca agreements). We thankfully acknowledge the computer
resources and technical support provided by these institutions.

\end{document}